\documentclass[conference,a4paper]{IEEEtran}
\usepackage{amsmath}

\ifCLASSOPTIONcompsoc
\usepackage[caption=false,font=normalsize,labelfont=sf,textfont=sf]{subfig}
\else
  \usepackage[caption=false,font=footnotesize]{subfig}
\fi

\usepackage{xcolor}
\usepackage{balance}

\ifCLASSINFOpdf
  \usepackage[pdftex]{graphicx}
\else
  \usepackage[dvips]{graphicx}
\fi
\ifCLASSOPTIONcompsoc
 \usepackage[caption=false,font=normalsize,labelfont=sf,textfont=sf]{subfig}
\else
 \usepackage[caption=false,font=footnotesize]{subfig}
\fi
\begin{document}
%
\title{RIS-Enhanced MIMO Channels in Urban Environments: Experimental Insights}


\author{\IEEEauthorblockN{
James Rains\IEEEauthorrefmark{1},
Anvar Tukmanov\IEEEauthorrefmark{2},
Qammer Abbasi\IEEEauthorrefmark{1}, 
\& Muhammad Imran\IEEEauthorrefmark{1}}
\IEEEauthorblockA{\IEEEauthorrefmark{1}
James Watt School of Engineering, University of Glasgow, Glasgow, UK, \\(james.rains, qammer.abbasi, muhammad.imran@glasgow.ac.uk)}

\IEEEauthorblockA{\IEEEauthorrefmark{2}
BT Labs, Adastral Park, Ipswich, UK, (anvar.tukmanov@bt.com)}
\thanks{Anvar Tukmanov acknowledges funding under UKRI Future Leaders Fellowship grant (MR/T019980/1)}
\thanks{James Rains' PhD is funded by EPSRC ICASE studentship (EP/V519686/1) with British Telecom}
\thanks{}}



\maketitle

\begin{abstract}
Can the smart radio environment paradigm measurably enhance the performance of contemporary urban macrocells? In this study, we explore the impact of reconfigurable intelligent surfaces (RISs) on a real-world sub-6 GHz MIMO channel. A rooftop-mounted macrocell antenna has been adapted to enable frequency domain channel measurements to be ascertained. A nature-inspired beam search algorithm has been employed to maximize channel gain at user positions, revealing a potential 50\% increase in channel capacity in certain circumstances. Analysis reveals, however, that the spatial characteristics of the channel can be adversely affected through the introduction of a RIS in these settings. The RIS prototype schematics, Gerber files, and source code have been made available to aid in future experimental efforts of the wireless research community.
\end{abstract}

\vskip0.5\baselineskip
\begin{IEEEkeywords}
Reconfigurable intelligent surfaces, MIMO, wireless communications, reconfigurable metasurfaces.
\end{IEEEkeywords}

%

\section{Introduction}
\IEEEPARstart{R}{econfigurable} intelligent surfaces (RISs) offer a means of controllability over radio environments that have until recently been assumed to be something our communication systems are completely at the mercy of. By embedding the capability to exert control over the wireless channel, we gain the power to reshape it according to our needs, transforming challenging propagation conditions into more favorable ones. Several recent works have sought to measure the performance benefits of including these smart surfaces within wireless channels at a broad range of frequencies currently in use by mobile broadband operators, as detailed in the review by Huang et al. \cite{Huang2022}. Sang et al. recently demonstrated the power of RISs to enhance the achievable data rate within a live 5G network \cite{Sang2022}. Most notably, the authors reported over a 100\% increase in throughput for users located within an indoor shadow zone. Another recent work experimentally demonstrated rank optimisation in RIS-assisted MIMO channels. Meng et al. \cite{Meng2023} developed an efficient optimisation algorithm to maximise the effective rank of a $2\times2$ MIMO link in a controlled indoor setting. The work presented here contains a set of measurements conducted within a real-world multiple-input multiple-output (MIMO) communication setup enhanced by a reconfigurable intelligent surface. We utilise a RIS beam search algorithm and demonstrate its efficacy over a $4 \times 4$ MIMO channel in an outdoor urban environment.

\section{Reconfigurable intelligent surface testbed}

\subsection{Hardware}

\begin{figure}[!t]
\centering
\includegraphics[width=0.9\columnwidth]{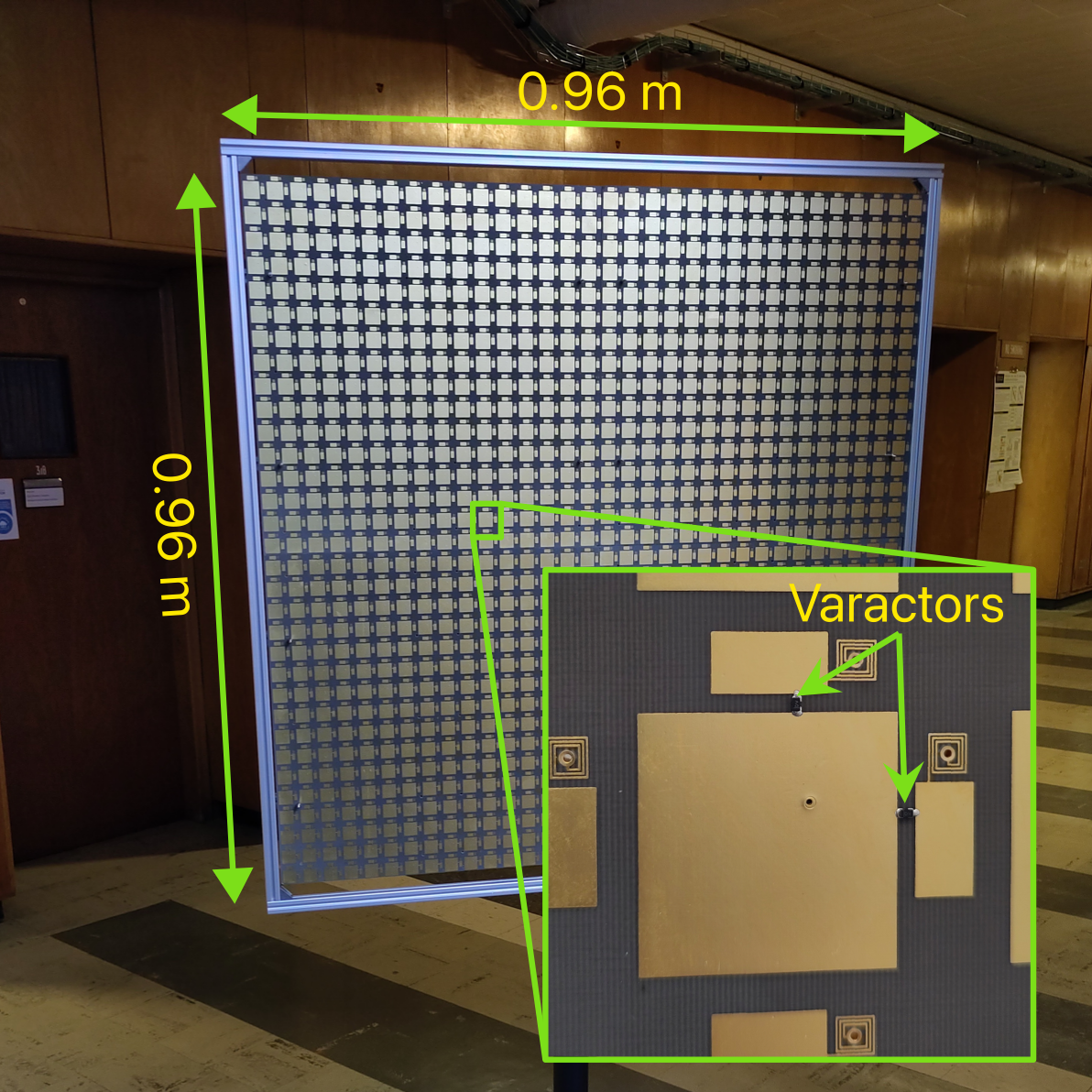}%
\caption{The RIS testbed employed in the measurements. Consists of 4 tiles with an arrangement of 16 $\times$ 16 unit cells each. Two varactor diodes are addressable on each unit cell, with two programmable bias voltage levels available to each diode. Schematics, Gerber files, and example code have been made available on Github \cite{Rains_OpenRIS_2022}.}
\label{ris_photo}
\end{figure}

The RIS we employ for these measurements is the varactor-based dual-polarised 1-bit low power RIS introduced in \cite{Rains2023a}, shown in Fig. \ref{ris_photo}. The RIS consists of 1024 elements in a 32 by 32 arrangement and is designed to operate within the 3.2 to 3.8 GHz 5G band. The elements are individually addressable in both polarisations. The lateral dimensions are 0.96 m by 0.96 m ($11\lambda_0 \times 11\lambda_0$). The tuneable elements and bias circuitry consume approximately 10 mW in their configured state and the testbed is powered by a portable battery bank.

\subsection{Beamsteering performance}

\begin{figure}[!t]
\centering
\subfloat[Vertical polarisation]{\includegraphics[width=0.9\columnwidth]{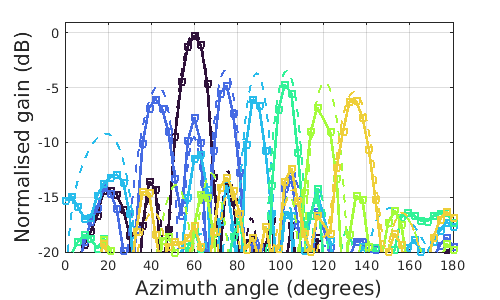}%
\label{fig_first_case}}

\subfloat[Horizontal polarisation]{\includegraphics[width=0.9\columnwidth]{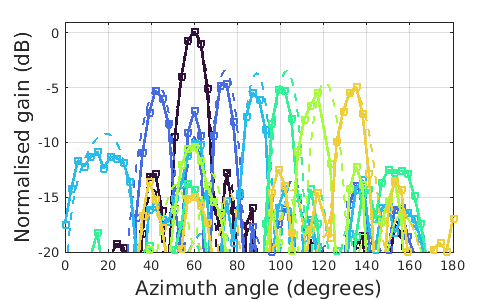}%
\label{fig_second_case}}
\caption{Measured (square curve) normalised gain patterns for $120^\circ$ far-field incidence on a single 16 $\times$ 16 RIS tile for vertical (a) and horizontal (b) antenna polarisations at 3.5 GHz. Showing beam patterns for maximum power between $45^\circ$ and $135^\circ$ in $15^\circ$ steps. Dashed curves are the approximate gain patterns using the model from \cite{DegliEsposti2022}.}
\label{beam_patterns}
\end{figure}

The far-field to far-field beamsteering performance for a single 0.48 m $\times$ 0.48 m RIS tile was ascertained for several configurations and incidence angles. These are based on the physics-compliant model introduced by Esposti et al. \cite{DegliEsposti2022}. Beam pattern measurements for $120^\circ$ incidence in azimuth (with $90^\circ$ being broadside to the RIS) are plotted in Fig. \ref{beam_patterns}(a) and (b) for vertical and horizontal antenna polarisation, respectively. It can be seen that the forward model provides a relatively reliable approximation for the performance of the main lobe in each configuration, although up to a 2 dB magnitude reduction can be noted for some of the non-specular configurations. We utilised this forward model within a particle swarm optimisation (PSO)-based beam search algorithm to determine any performance benefits within an urban macrocell.  

%

\section{Beam Search Algorithm}

During initial access by user equipment (UE) in 5G NR networks, beam sweeping is typically employed. The 5G base station, or gNodeB (gNB), performs a beam sweep and the user reports its preferred beam which is then set by the gNB. Protocols are in place for continued beam reporting and switching according to the varying channel conditions. Inspired by this simple approach to beam selection, we aim here to determine the utility of a beam search algorithm facilitated by the RIS. When there's an obstruction between the transmitter and receiver, a properly positioned RIS can create a virtual line-of-sight (VLoS) channel. In cases of significant blockage, this VLoS channel becomes the dominant propagation mechanism, potentially allowing us to use the same model for generating beam patterns such as those in Fig. \ref{beam_patterns} for link enhancement. Assuming we know the position of the transmitter, we limit the unknown parameters to the spherical coordinates of the receiver, $r_{rx}$, $\theta_{rx}$, $\phi_{rx}$. Additionally, a fourth parameter is introduced which relates to a $180^\circ$ phase shift in the main lobe between two similar configurations. This effectively amounts to flipping all of the bits of a given configuration with the aim of facilitating constructive interference of a single in the VLoS path with additional significant NLoS components. A PSO algorithm with these 4 parameters was employed with the goal of maximising the channel gain of a $4\times4$ MIMO system. 


\section{Experiment setup}

\begin{figure}[!t]
\centering
\includegraphics[width=0.83\columnwidth]{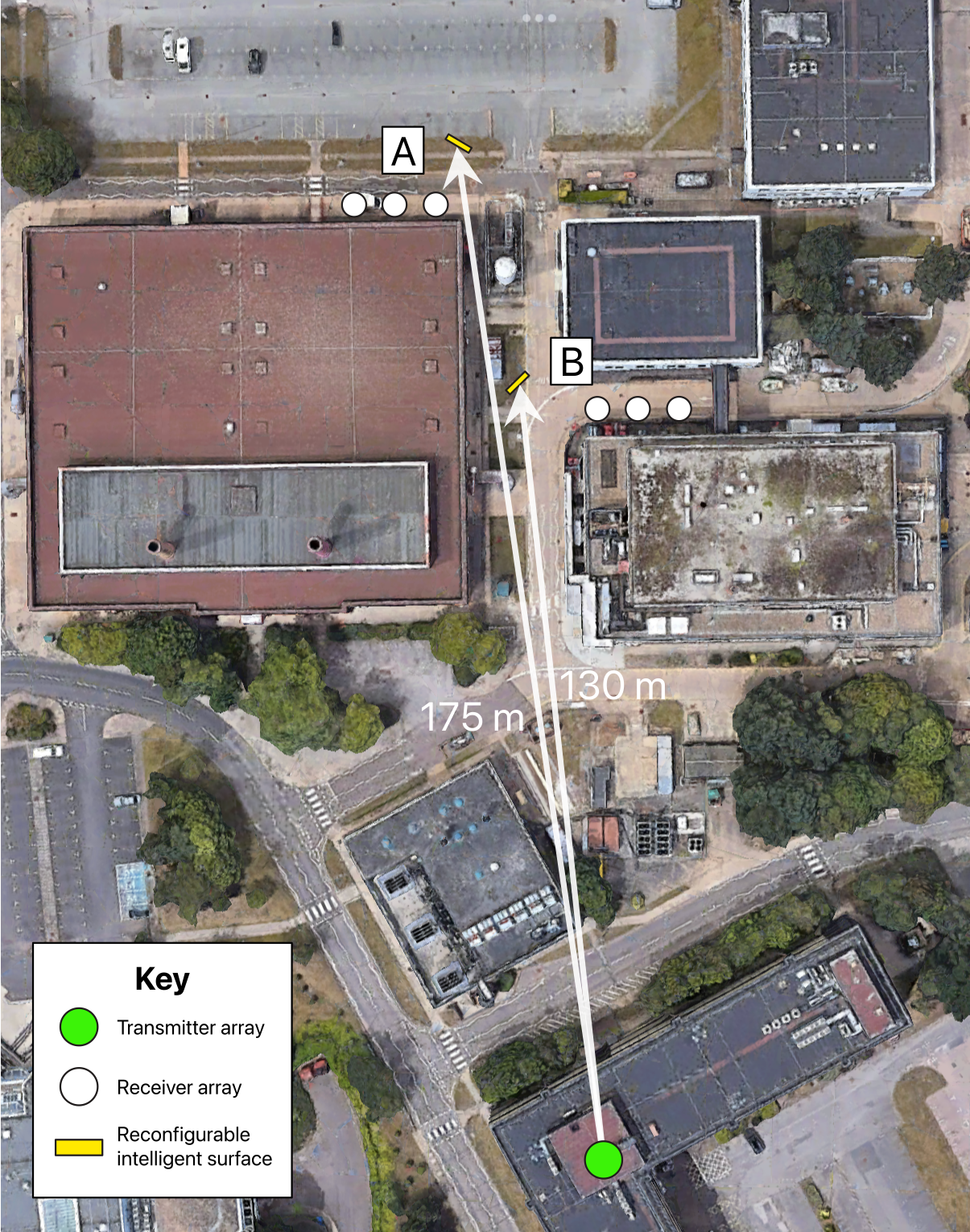}%
\caption{Experiment setup at Adastral Park, Ipswich, UK. A rooftop base station antenna is directed northwards towards a RIS and receivers placed at street level. The two measurement locations are denoted A and B.}
\label{exp_setup}
\end{figure}

\begin{figure}[!t]
\centering
\includegraphics[width=0.83\columnwidth]{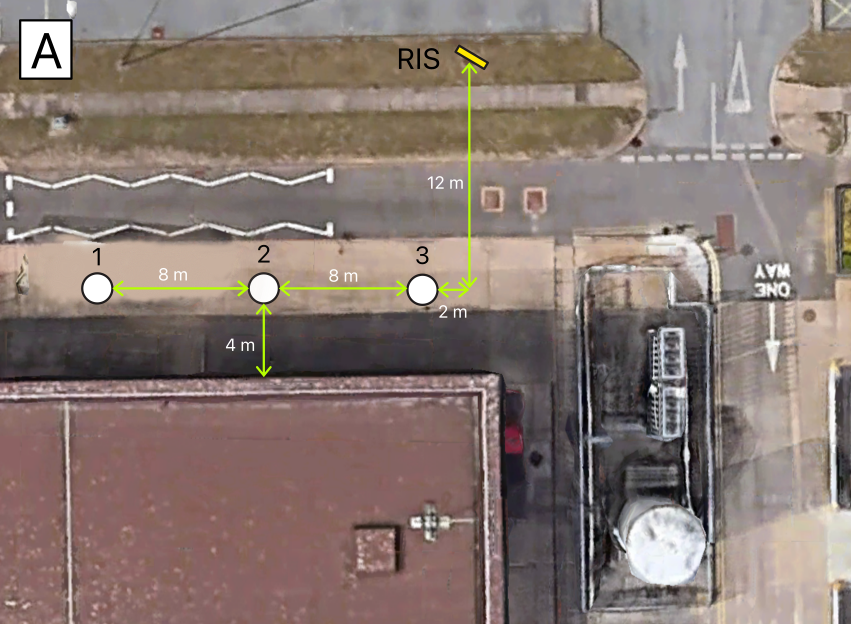}%

\caption{Measurement zone A showing the RIS position and 3 receiver locations.}
\label{meas_zones}
\end{figure}

The experiment was undertaken at BT Labs at Adastral Park, situated in the east of England. The base station (transmitter) antenna array position, the RIS, and the receiver array positions are depicted in Fig. \ref{exp_setup}. Two measurement locations, labeled A and B, were chosen because they have occlusions from surrounding buildings between the receiver and transmitter array. The distances between the RIS and the transmitter antenna at zones A and B are 175 m and 130 m, respectively. Diagrams of zones A and B are plotted in Fig. \ref{meas_zones}(a) and (b), respectively. For brevity, a subset of the results for Zone A are covered in this paper.
\subsubsection{Transmitter array} The transmitter antenna is a Commscope RRZZHHTTS4-65B-R7 multiband sector antenna situated on a rooftop, at a height of approximately 25 m from ground level. This antenna ordinarily serves as part of a macrocell for the site. The 8T8R 3300-3800 MHz section of the antenna consists of 4 columns of $\pm45^\circ$ polarised antenna elements. Each column is driven by two ports, with each corresponding to the respective polarisations, for a total of 8 input ports. In this work, we have utilised the center 4 ports of the array, corresponding to two sets of $\pm45^\circ$ polarised array columns. Each array column provides a 16 dBi gain with beamwidths of $89^\circ$ and $6.5^\circ$ in azimuth and elevation, respectively. The antenna is mechanically and electrically tilted such that the main lobes points toward $0^\circ$ in elevation.

\subsubsection{Receiver array} The majority of published field trials tend to filter out potentially beneficial multipath components through use of highly directional horn antennas. This does not provide a fair reflection of the performance of mobile networks at sub-6 GHz, where handsets typically employ electrically small antennas that are omnidirectional in nature. Instead, it sets the baseline performance unrealistically low, making the introduction of the RIS appear to significantly benefit the network more than it would in a real-world implementation. For a more realistic analysis, the receiver antenna employed here is a 4-element ground-backed crossed-dipole array. This was selected as a compromise between ascertaining realistic channel measurements and minimising perturbations by measurement equipment and personnel in proximity to the antenna.

\begin{figure}[!t]
\centering
\includegraphics[width=0.65\columnwidth]{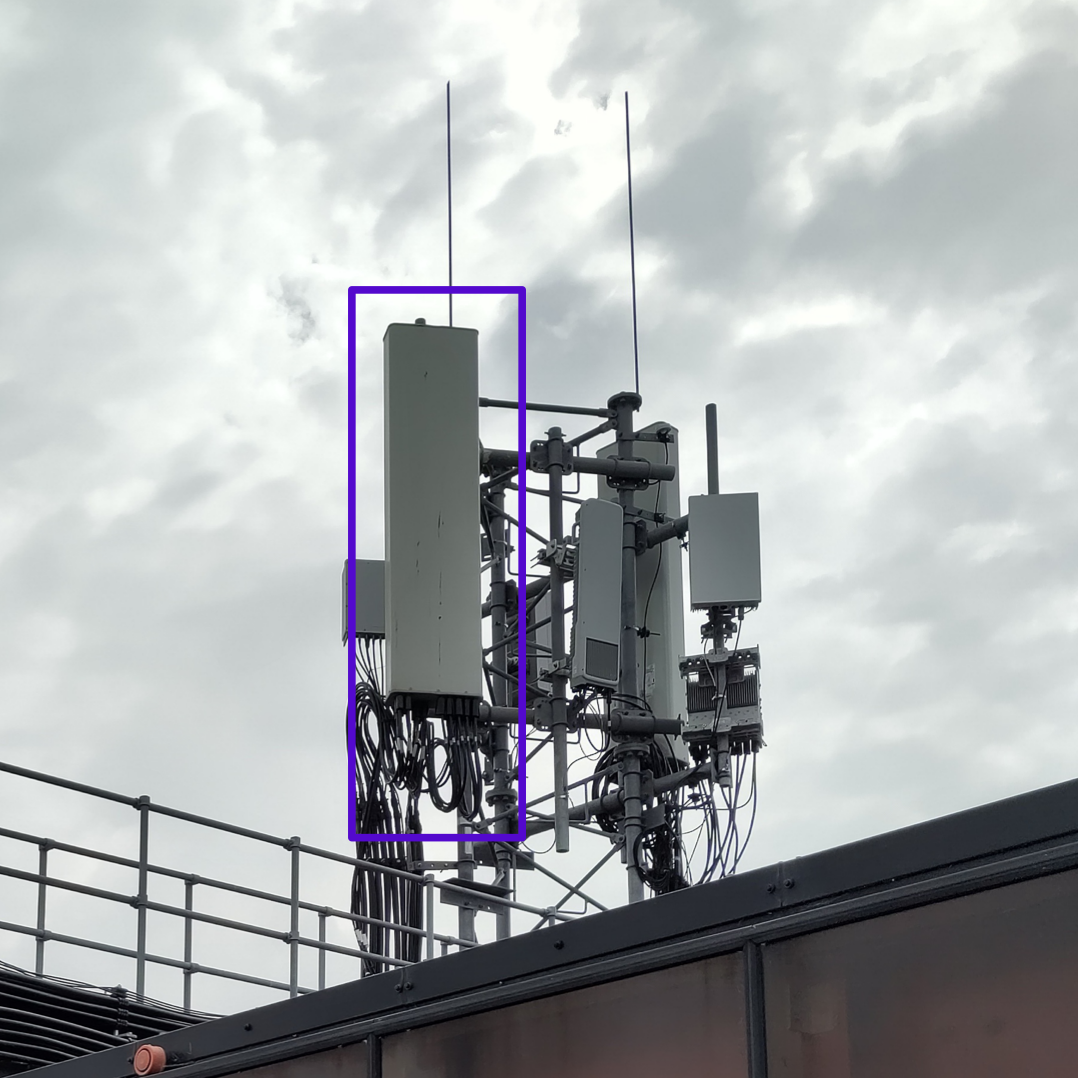}%
\caption{Sector antenna used in the measurements. The main lobe is directed north, with $90^\circ$ and $6.5^\circ$ beamwidths in azimuth and elevation, respectively.}
\label{commscope}
\end{figure}

\begin{figure}[!t]
\centering
\includegraphics[width=0.5\columnwidth]{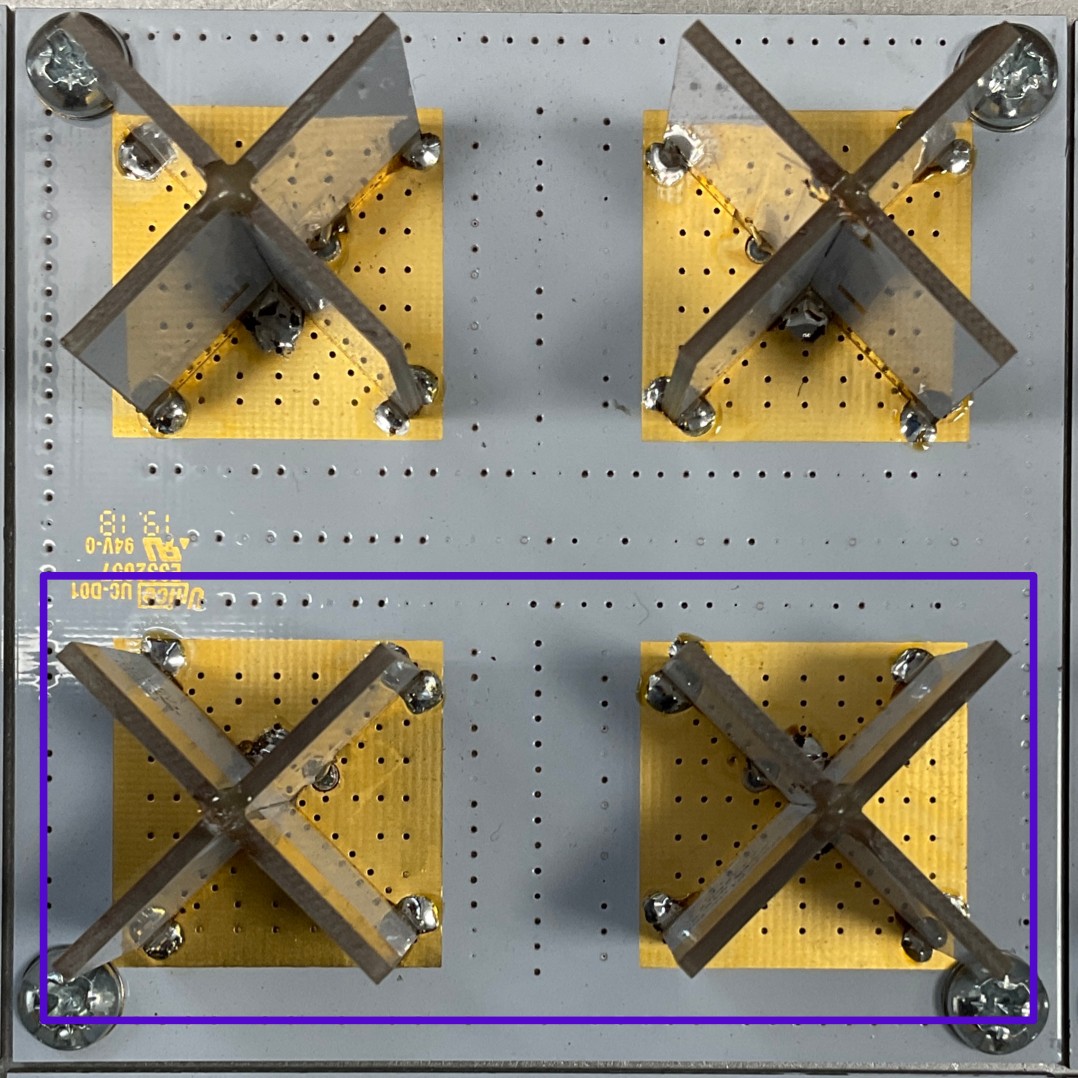}%
\caption{Broadband dipole elements used for the receiver. Highlighted elements are addressed while remaining elements are terminated in matched loads.}
\label{dipole}
\end{figure}

\subsubsection{Measurement system} 
\begin{figure}[!t]
\centering
\subfloat[Rig module]{\includegraphics[width=0.9\columnwidth]{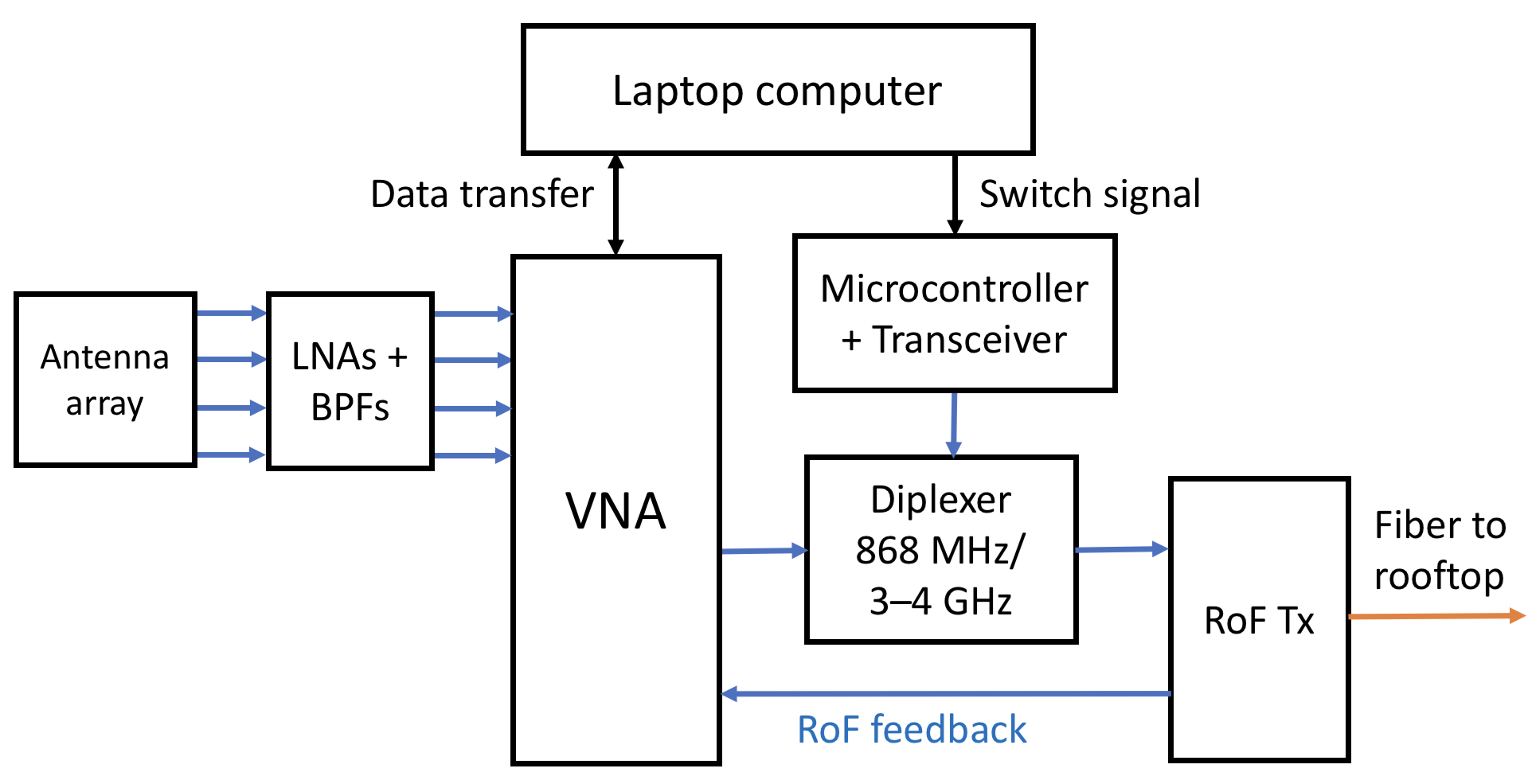}%
\label{fig_first_case}}

\subfloat[Rooftop module]{\includegraphics[width=0.9\columnwidth]{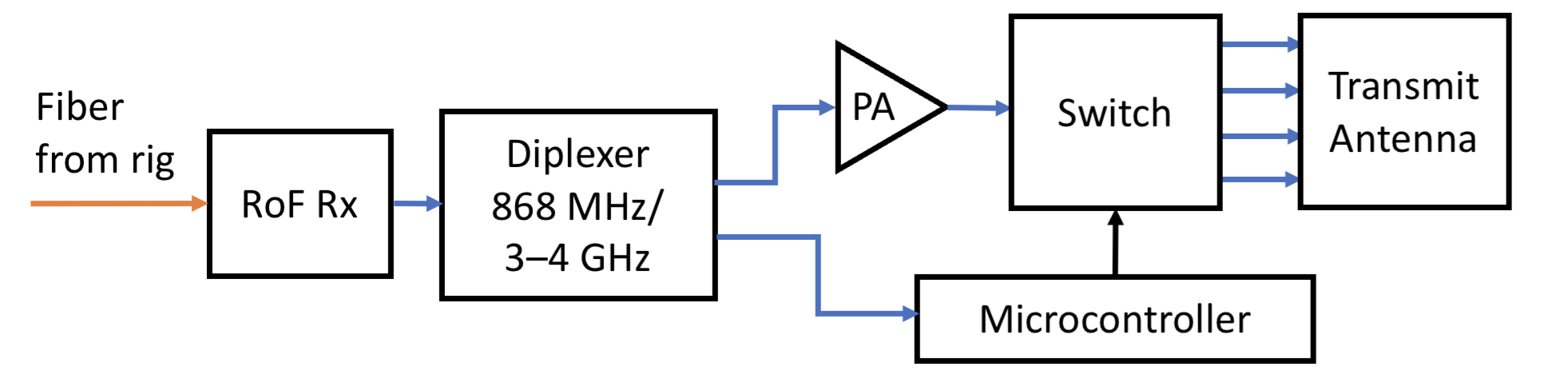}%
\label{fig_second_case}}
\caption{System diagram of the receiver rig (a) and transmitter setup (b) for the long-range VNA-based channel sounder.}
\label{rig_roof_diagrams}
\end{figure}

The measurement system is centered around a Rohde and Schwarz ZNBT 12-port vector network analyser (VNA) located on a rig at the receiver location. A system diagram of the rig is shown in Fig. \ref{rig_roof_diagrams}(a). To form a closed-loop measurement system over such a long distance, thus retaining accurate phase information, we extended the range of the VNA through an RFOptic radio over fiber (RoF) transmitter and receiver pair. This was made feasible due to existing optical fiber infrastructure at the measurement location. A sounding signal is sent from one of the VNA ports to a RoF transmitter module which modulates a laser. Additionally, a switch control signal line is added through a diplexer. These signals are optically transmitted across the site to a room directly beneath the base station antenna, where they are detected and transformed back into electrical signals by a RoF receiver, as shown in Fig. \ref{rig_roof_diagrams}(b). Part of the received optical signal is sent back down the fiber and detected at the rig in order to be used as a reference to calibrate out the delay introduced by the fiber. Following the RoF receiver is a diplexer used to separate the switch control signal path from the sounding signal path. The sounding signal is fed into a 16 W Minicircuits ZHL-16W-43-S+ power amplifier (PA), followed by a SP4T mechanical switch whose 4 output ports are connected to 4 ports of the antenna array via low-loss cables. The switch control signal is received by a transceiver connected to a microcontroller, which in turn sets the mechanical switch states, addressing the respective transmitter antenna ports. To close the loop, the 4 ports of the receiver array are each connected to low noise amplifiers (LNAs) followed by bandpass filters (BPFs). The receiver antennas are connected to four ports on the VNA, allowing concurrent measurements of the received signals from the receiver array elements.

\section{Optimisation performance}

\subsection{Measurement procedure}

\begin{figure}[!t]
\centering
\subfloat[Position 1]{\includegraphics[width=1\columnwidth]{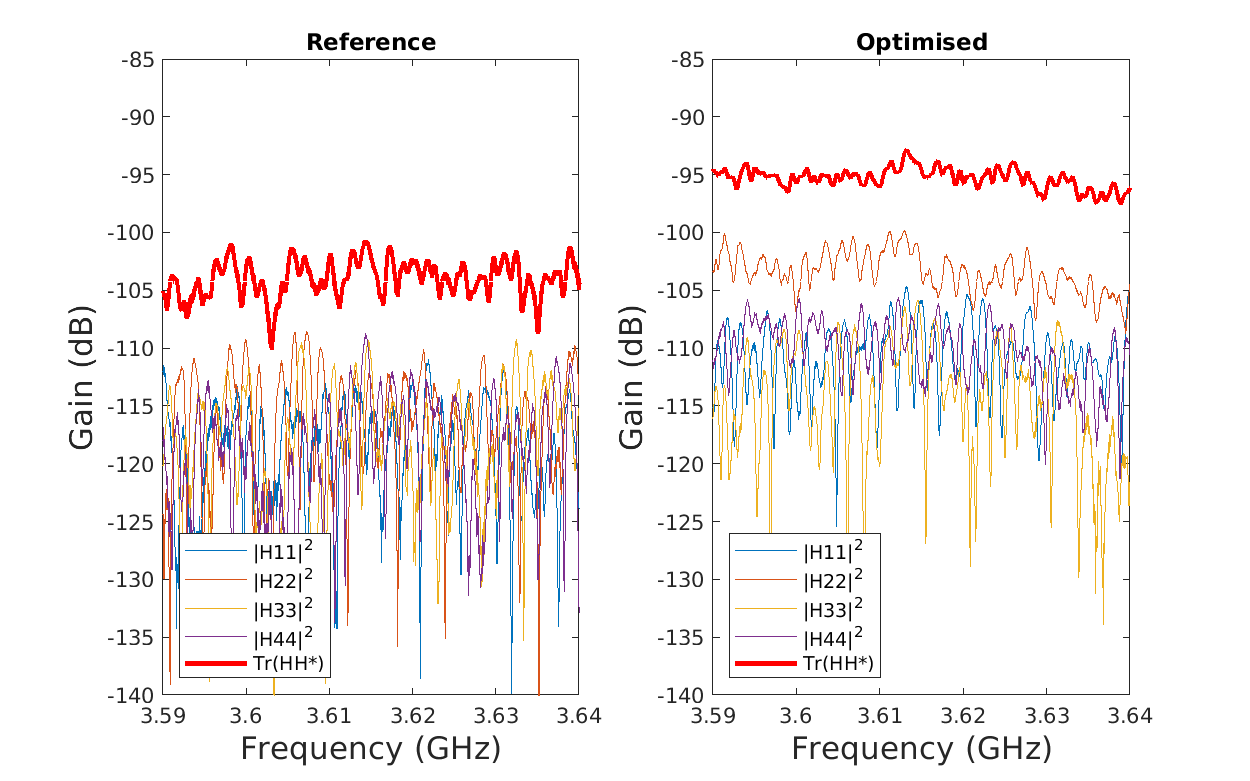}%
\label{fig_first_case}}


\subfloat[Position 3]{\includegraphics[width=1\columnwidth]{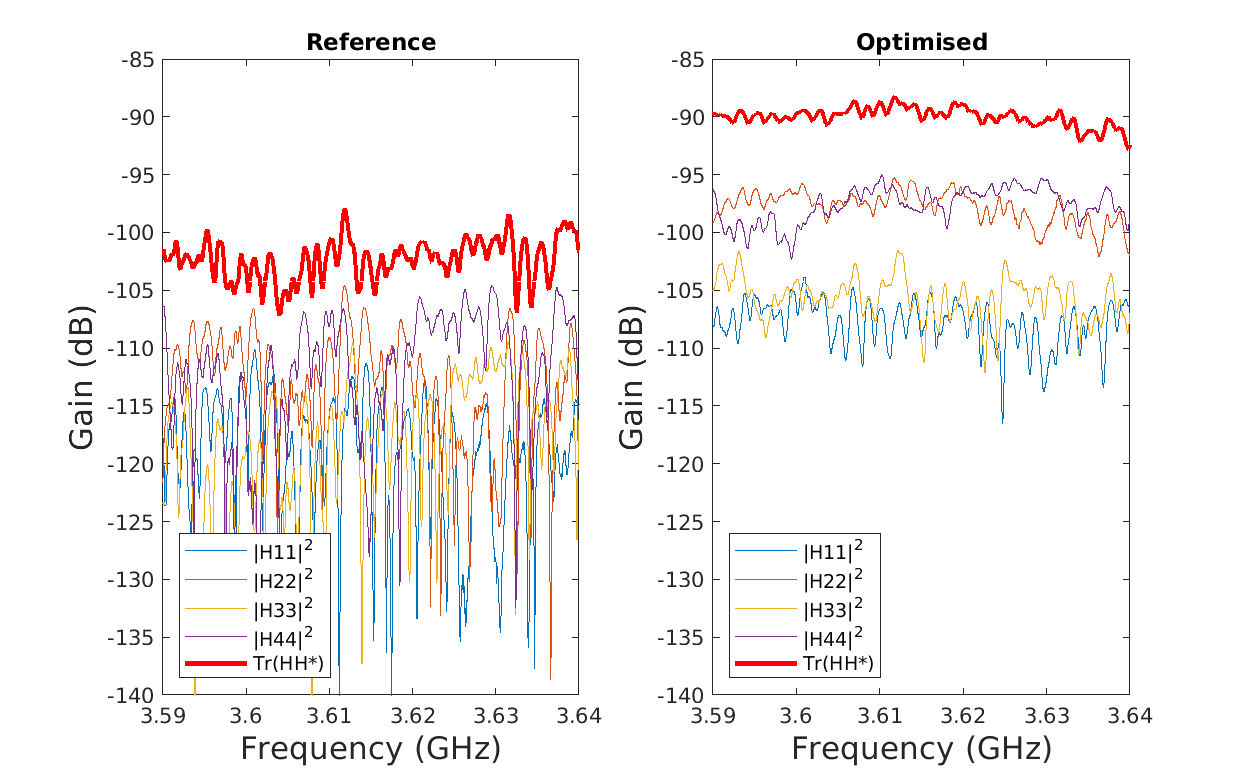}%
\label{fig_third_case}}

\caption{Selected channel magnitudes, $H_{11}$ to $H_{44}$, and MIMO channel gain for the 3.59 to 3.64 GHz band in zone A. Reference and optimised cases at positions 1 and 3, (a) and (b), respectively.} 
\label{chan_mags}
\end{figure}

For each receiver location, the PSO-based beam search algorithm was employed to find the configuration which maximised the channel gain for a 50 MHz band. The RIS was tuned to operate at 3.5 GHz. There was an active 5G NR cell on the site operating between 5.53 and 5.58 GHz. Therefore, two 50 MHz bands operating above and below this range were selected in order to avoid unnecessary interference. Namely, 3.47 to 3.52 GHz and 3.59 to 3.64 GHz. Channel measurements were taken before and after the introduction of the RIS. These consist of 16 frequency domain sweeps corresponding to the complex transmission parameters between each transmitter and receiver. Additionally, a feedback measurement from the RoF section is taken to compensate for the fiber delay in post-processing. A subset of the channel parameters after de-embedding are plotted in \ref{chan_mags}(a) to (c) for locations 1 to 3, respectively. Additionally, the total power gain for the complete $4\times4$ MIMO channel matrix is plotted according to equation (\ref{eqn:chan_gain}):

\begin{equation}
G = \text{Tr}\left(\mathbf{HH}^H\right)
\label{eqn:chan_gain}
\end{equation}

This is the gain of the MIMO channel assuming energy is spread equally between all four transceiver ports \cite{Tse2005}. 

\subsection{Zone A}

\begin{figure}[!t]
\centering
\includegraphics[width=0.9\columnwidth]{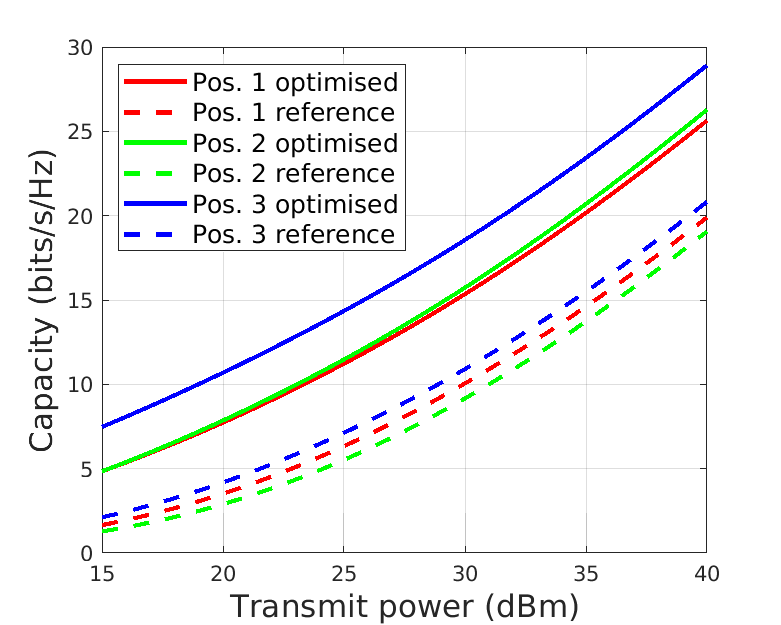}%
\caption{Predicted channel capacity for zone A between 3.59 and 3.64 GHz.}
\label{cap_v_txp}
\end{figure}

The RIS in Zone A was oriented so that its broadside direction was set at a $30^\circ$ clockwise angle relative to the line of receiver locations. This angle was chosen to ensure a sufficiently large capture area for incident radiation while also reducing the obliqueness of the angle presented to Location 1. The maximum channel capacity is found by applying the water-filling principle on the transmit covariance matrix. Predicted maximum channel capacity versus transmit power for the 3 measurement positions for the reference and optimised cases are plotted for the 3.59 to 3.64 GHz band. As a tangible example, in the UK, a typical EIRP limit in active antenna systems in the sub-6 GHz bands is 44 dBm/5 MHz per sector. For a 50 MHz channel bandwidth, this results in a maximum EIRP of 54 dBm. Given the base station antenna gain of 20.4 dBi, this results in a maximum transmission power of 33.6 dBm. Referring to Fig. \ref{cap_v_txp}, for positions 1 to 3, the reference case (i.e., before the RIS introduction) can theoretically facilitate transmission rates of up to 0.635, 0.590, and 0.687 Gbps, respectively. With the introduction of the RIS, these values are increased to 0.908, 0.933, 1.07 Gbps, respectively, thereby approaching and even exceeding the maximum downlink data rates of 1 Gbps at 5G mid-band.


\begin{figure}[!t]
\centering
\includegraphics[width=0.9\columnwidth]{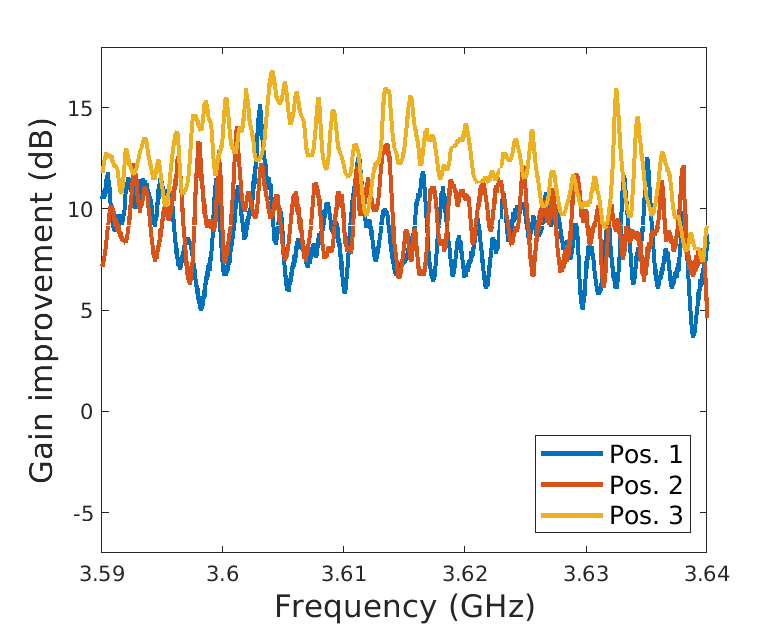}%
\caption{MIMO channel gain improvement for the 3.59 to 3.64 GHz band.}
\label{gain_comparison}
\end{figure}

\section{Effects on spatial structure}

\begin{figure}
\centering
\includegraphics[width=0.9\columnwidth]{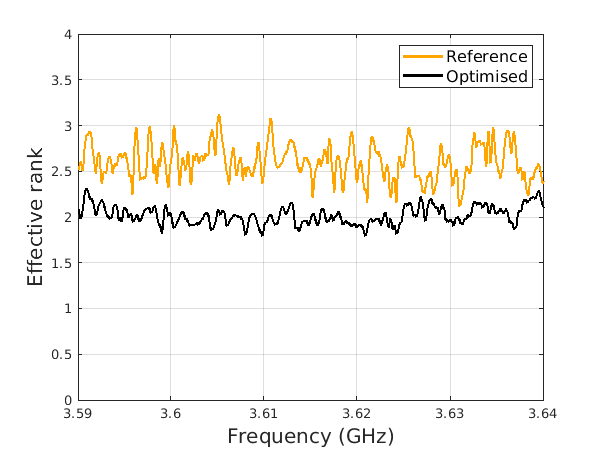}%
\caption{Effective rank for position 3, zone A within the 3.59 to 3.64 GHz band.}
\label{fig:effective_rank}
\end{figure}

The primary advantage of MIMO systems lies in their ability to enable concurrent communication streams through the utilization of orthogonal subchannels. One measure of the channel dimensionality is the \textit{effective rank} introduced by Roy and Vetterli \cite{7098875}. The effective rank quantifies the number of independent spatial dimensions effectively utilised by the channel, determined by the effective number of non-negligible singular values. For the $4\times4$ MIMO system employed here, a maximum effective rank of 4 is possible. In practice, the effective rank can vary profusely and is largely dependent on the richness of the multipath and mutual coupling between the antennas. This has been plotted for position 3, Zone A, in Fig.  \ref{fig:effective_rank}. The introduction of the RIS has detrimental implications for the spatial structure of the channel in this instance, with an effective rank reduction across the band. The mean effective ranks for positions 1 to 3, Zone A, for the reference case were calculated as 2.74, 2.87, and 2.61, respectively. With the introduction of the RIS, these are reduced to 2.24, 2.43, and 2.02, respectively. In this implementation, the increase in channel gain clearly comes at the cost of reduced diversity. This effective rank reduction could, for example, result in the communication system being more sensitive to interference \cite{Tse2005}.

\section{Conclusion}

In this paper, we have presented a snapshot of the results obtained from a comprehensive measurement campaign on RIS-aided MIMO communications within existing 5G network infrastructure. Our analysis of these results has revealed a channel gain enhancement of 10 to 15 dB under specific conditions, achieved through the deployment of a RIS with a low-complexity beam search algorithm. The introduction of RISs into currently deployed 5G networks holds the promise of significantly improving network performance by addressing underperforming regions within cells. RIS prototype information is available at http://www.github.com/jimrains/OpenRIS.

\section*{Acknowledgment}

The authors would like to extend our gratitude to the WiDeS group at USC, as well as the Wireless Research Team, Fundamental Research Team, and Kristan Farrow at BT for their support and guidance during these measurements. 

\bibliographystyle{ieeetr}
\bibliography{EuCAP2024_template}

\begin{thebibliography}{1}

\bibitem{Huang2022}
J.~Huang, C.-X. Wang, Y.~Sun, R.~Feng, J.~Huang, B.~Guo, Z.~Zhong, and T.~J. Cui, ``Reconfigurable intelligent surfaces: Channel characterization and modeling,'' {\em Proceedings of the {IEEE}}, vol.~110, pp.~1290--1311, Sept. 2022.

\bibitem{Sang2022}
J.~Sang, Y.~Yuan, W.~Tang, Y.~Li, X.~Li, S.~Jin, Q.~Cheng, and T.~J. Cui, ``Coverage enhancement by deploying {RIS} in 5g commercial mobile networks: Field trials,'' {\em {IEEE} Wireless Communications}, pp.~1--21, 2022.

\bibitem{Meng2023}
S.~Meng, W.~Tang, W.~Chen, J.~Lan, Q.~Y. Zhou, Y.~Han, X.~Li, and S.~Jin, ``Rank optimization for mimo systems with ris: Simulation and measurement,'' July 2023.

\bibitem{Rains2023a}
J.~Rains, J.~U.~R. Kazim, A.~Tukmanov, L.~Zhang, Q.~Abbasi, and M.~Imran, ``Fully-addressable varactor-based reflecting metasurface with dual-linear polarisation for low power reconfigurable intelligent surfaces,'' in {\em 2023 17th European Conference on Antennas and Propagation ({EuCAP})}, {IEEE}, Mar. 2023.

\bibitem{Rains_OpenRIS_2022}
J.~Rains, ``Github: Openris http://www.github.com/jimrains/openris,'' Dec. 2022.

\bibitem{DegliEsposti2022}
V.~Degli-Esposti, E.~M. Vitucci, M.~D. Renzo, and S.~A. Tretyakov, ``Reradiation and scattering from a reconfigurable intelligent surface: A general macroscopic model,'' {\em {IEEE} Transactions on Antennas and Propagation}, vol.~70, pp.~8691--8706, Oct. 2022.

\bibitem{Tse2005}
D.~Tse and P.~Viswanath, {\em Fundamentals of Wireless Communication}.
\newblock Cambridge University Press, May 2005.

\bibitem{7098875}
O.~Roy and M.~Vetterli, ``The effective rank: A measure of effective dimensionality,'' in {\em 2007 15th European Signal Processing Conference}, pp.~606--610, 2007.

\end{thebibliography}

\end{document}